# On dynamical systems theory in quantitative psychology and cognition science: A fair discrimination between deterministic and statistical counterparts is required

**Adam Gadomski**

*Group of Modeling of Physicochemical Processes, Institute of Mathematics & Physics,*

*University of Science & Technology, PL-85796 Bydgoszcz, Poland; agad@utp.edu.pl*

**Marcel Ausloos**

*GRAPES, B-4031 Liège, Federation Wallonie-Bruxelles, Belgium; marcel.ausloos@ulg.ac.be*

*School of Business, University of Leicester, LE1 7RH Leicester, United Kingdom; ma683@le.ac.uk*

**Tahlia Casey**

*Department of Biology, Delaware State University, 1200 N DuPont Highway, Dover, DE 1990, USA; tahlia.c.casey@gmail.com*

Correspondence should be addressed to: Adam Gadomski, University of Science & Technology, al. Kaliskiego 7/421, PL-85796 Bydgoszcz, Poland

*Abstract: The present communication addresses a set of observations, obeying both deterministic as well as statistical formal requirements, and serving to operate within the framework of the dynamical systems theory, with a certain emphasis placed on initial data. It is argued that statistical approaches can manifest themselves nonunivocally, leading to certain virtual discrepancies in psychological and/or cognitive data analyses, termed sometimes in literature as, questionable research practices. This communication points to the demand for a deep awareness of the data origins, which can indicate whether the exponential (Malthus type) or the algebraic (Pareto type) statistical distribution ought to be effectively considered in practical interpretation. This is also related to the question of how frequently patients behave in a specific way, and the significance of these behaviors in determining a patient's progression or regression, involving a certain memory effect. In this perspective, it is discussed how a sensitively applied hazardous or triggering factor can be helpful for well-controlled psychological strategic treatments, also those attributable to obsessive-compulsive disorders or even self-injurious behaviors, with their both criticality- and complexity-exploiting relations between a therapist and a patient.*

## INTRODUCTION

In the following communication, we wish to add to Gelfand and Engelhart (2012), some emphasis on the dynamical systems theory (DST) applications in practical and theoretical psychology, and cognition science, addressing the role of short and long term memory effects, including diffusional aspects, as well as the distinction to be made between continuous time approximations and discrete time measures - the latter leading to chaotic behavior perspectives (Barton, 1994; Mandel, 1995; Guastello and Gregson, 2011).

A major characteristic feature of any dynamical system concerns the values, of course representative of its descriptive variables, when taken at one given time, are regarded as functions of those same variables at earlier times, whatever their measured values. In a recent opinion, Gelfand and Engelhart (2012), emphasized the importance of familiarizing the lay reader, or traditional psychologists, with DST. To be concise, this area of mathematical physics is used to describe the dynamics of complex systems through differential equations that are often nonlinear. The DST has been applied outside of physics, with definite noticeable success, as in neuroscience and cognitive development, as seen in the continued use of the Neo-Piagetian theories of cognitive development, see e.g. Izhikevich (2007).

Although the pure rationale of bearing upon DST in neuroscience and cognition research seems unquestionable, the way in which Gelfand and Engelhart (2012) demonstrated the problem's significance is indeed worthy of re-examination in terms of specific examples, - in order to make the perspective and research aims sufficiently appealing. Moreover, in what follows, we demonstrate how DST ideas and techniques can be performed based on the Malthusian type of reasoning. Let it be recalled that a Malthus (1798) idea/reasoning implies the fear of an exponential (population) growth. For defusing the critical aspect, Verhulst (1845, 1847) constrained the population growth by a "capacity" of the area capable of sustaining the population. Neuroscientists and applied psychologists have adapted the ideas of Verhulst's about the capacity of an area into the "capacity of memory" in order to apply constraints to the use of DST for predicting and controlling behavior (Ausloos and Bromberg, 1972).

# DETERMINISTIC APPROACH AND INITIAL DATA

In (Gelfand and Engelhart, 2012) a very simple discrete DST example is invoked, namely the basic one-step map: $x(t+1) = ax(t)$, with a two-fold ("amphiphatic") variable (irrespective of whether stochasticity is present or not), $x$, playing a doubled role: at instant $t$, $x$ is an independent variable, while at another instant, $t+1$, $x$ is a dependent one; notice that $a$ stands for a (positively valued) growth coefficient. Conversely, if $a$ is negative, it is a dissolution/depression coefficient. This DST mathematical map is a uniquely stated feedback system description, such that if one inputs $x(t)$, then, after multiplication by $a$, one receives an output of $x(t+1)$.

The map presented, however, differs for instance, from the Fibonacci type, $x(t+2) = x(t+1) + x(t)$, wherein <u>two</u> input data sets are required before getting an output, from consecutive times strictly prior to the observation time. For example, starting from *x(0)=x(1)=1*, following the rule $x(t+2) = x(t+1) + x(t)$, one yields the well-known numbers: $2, 3, 5, 8, 13, 21, ...$, i.e., being unevenly distributed, - called a Fibonacci series. The sequence is well known in natural systems (Rehmeyer, 2007), and has been connected to memory aspects (Baddeley, 1992). The fundamental distinction between the simplest DST (Gelfand and Engelhart, 2012), and the Fibonacci map, describing deterministic branching (or, splitting) processes, involving, e.g. Bethe lattice (Gadomski and Ausloos, 2006), looks profound indeed, and has yielded an added value for quantitative psychological methodology of measurements. Namely, let us repeat, the latter map needs two input (one-and-one) data to get the next dependent variable value, whereas the former map needs only a single input.

From a simple psychological viewpoint, the use of one input data point would be more appropriate to mimic amnesic syndroms or memory losses (Lejeune, 1998), or could also be loosely named, a Markov process (Norris, 1998), or at most, a very short-range memory by a person under treatment (Baddeley, 1992). The use of two inputs, in turn, looks more predisposed to hinder (longer range) memory, involving human aims, quite in terms of stochastic processes, and non-Markovian systems' dynamics, if it is (often justly) taken that any $x(t)$ is a stochastic, thus, statistical and time-dependent variable (Fuliński et al., 1998).

In the following, a proper distinction between deterministic and statistical approaches, is worth further argument. First, putting things as simply as they can be treated, let it be realized that the map $x(t+1) = ax(t)$, see Gelfand and Engelhart (2012), is fully equivalent to $[x(t+1) - x(t)]/x(t) = a - 1$, taken with $0 < a < 1$, - upon slightly confining the generality of the presented argumentation to a fractional prefactor included in the (discrete) map. Replacing for the moment the difference $x(t+1) - x(t)$ by its (continuous) infinitesimal increment $dx$, and putting $a = 1 - dt$, with its respective, very small time increment $0 < dt \ll 1$, one provides via $dx/x = -dt$ the exponential, nonlinear solution, $x(t) = \exp(-t)$, obeying, however, the discrete rule of $x(t+1) = (1/e)x(t)$ when $e \simeq 2.71$. Let us point out that DST, indeed, can be seen either as continous time dynamic processes or through recurrence, discrete time equations. Exponential functions in psychopathology have also been studied specifically by Bond and Guastello (2013). They have found that dynamical disease is one in which the symptoms appear and disappear (like balloon air emptying and filling in, see below) over time in a deterministic way, resembling a chaotic pattern. By determining the dynamic structure of the temporal pattern it could be possible to gain at least some insight into the triggers for symptoms. It has been argued thouroughly whether obsessive-compulsive disorder (OCD) could be treated as a dynamical disease because of the intermittent outbursts of ritual behaviors. Specifically, one-week diaries were collected from about twenty clinical cases with almost the same number matched controls to record both the occurrence of rituals as they transpired over time and the influence the family may have expressed upon the spatiotemporal structure of symptoms. Comparisons of nonlinear regression parameters and Lyapunov exponents, exhibiting exponentially fast divergence of the corresponding dynamic trajectories, revealed that OCD developed a low-dimensional deterministic structure. A nonlinear model explained more than ten times greater the variance as compared to its linear counterpart. Family reactions and emotional responses accounted for only a small increase in the variance explained by the nonlinear regression model or in the amount of turbulence. Family reactions and emotional responses do not influence too much towards making the rituals go away, but instead may amplify the dynamics. In addition, significant rank order correlations were disclosed between both variances for the time series for each logbook and Lyapunov exponents with symptom severity and family reactions, cf. Bond and Guastello (2013).

Continuing the reasoning, let us ascertain that the solution of the single input map is, of course, within the accuracy of setting the integration constant equal to one, $\exp(-t)$, i.e. a decreasing function of time. For example, it is known to describe the simple dynamic physical process of air emptying a spherical balloon of radius $R$ by means of a constant depressurization factor $\Delta\pi < 0$ such that $dR/R \propto \Delta\pi < 0$ applies, wherein $dR$ denotes a very small change in the balloon radius value. An analogy can be made to purposefully using triggering language consistently to excite or irritate a subject in order to stimulate an emotional response, which will enable the therapist to change and quantitatively observe the patient's emotions. In this process one is able to utilize the environment as a tool to modify the directional behavior change towards positive behavioral activity. It is to be proclaimed here through analogy arguments that the deterministic dynamical picture of balloon collapsing can coexist naturally with certain psychological issues, as those related to depression. These may emerge when an individual is able to openly express her/his emotions, after removing sources of criticism, and/or eliminating other significant stressors of similar kinds, allowing the subject to return to her or his positive behavioral activity as, for example, discussed by Osborne (2010).

## STOCHASTIC (STATISTICAL) APPROACH AND EMERGENCE OF MEMORY EFFECT VIEWED IN TERMS OF REDUCED VARIANCE

Going a step further, if one realizes that as earlier discussed in holographic memory and cognitive models by Ausloos and Bromberg (1972), or by extension schizophrenia or neurodegenerative Alzheimer-type disorders, even in their infancy stages, due to the loss of short (phonological) memory experienced by patients (Baddeley, 1992), there must be always an inevitable diffusion (or, fuzzy) aspect of memory capacity. This points to the observation that approaches addressing diffusion and non-feedback must be utilized in describing such psychological phenomena of interest. Although the creation of dissipative structures in behavioral neuroscience is fascinating and greatly interesting (Kennedy and Raz, 2009; Heinke and Mavritsaki, 2009; Schiepek et al. (2015)), without digressing further in complex diffusion approaches, let it be recalled that the diffusion coefficient is a measure of fluctuations' correlations in Wiener, so-called memoryless information, processes (Fuliński et al., 1998, and refs. therein; Bazzani et al., 2003; Ausloos, 2015).

As concerns the interaction of a system (patient) with her/his (possibly fuzzy) surroundings, it is worth pointing out that mental illnesses are defined as dysfunctions of the system-environment-fit. Individual symptoms are viewed as the result of illness-inflicting and illness-maintaining relationship patterns within the context of the patient's loved ones and significant others. The patient's significant others are often included in the therapy process, Schiepek et al. (2015); Schiepek et al. (2016) what, however, augments sometimes markedly the number of virtual influencial cofactors when examining the system as a whole. There is no doubt that psychotherapy is a dynamic process produced by a complex system of interacting variables. Even though there are qualitative models of such systems the link between structure and function, between network and network dynamics is still lacking. In what follows a reasoning is offered to realize these links. The proposed model, cf. Schiepek et al. (2016), is composed of five state variables: problem severity, success and therapeutic progress, motivation to change, emotions, insight, and new perspectives, interconnected by an adequate number of functions. The shape of each function is modified by four "social" parameters: capability to form a trustful working alliance, mentalization and emotion regulation, behavioral resources and skills, self-efficacy, and reward expectation. To put it sociotechnically, the parameters play the role of competencies, which translate into the concept of control parameters in synergetics (Schiepek et al., 2015), or when employing the notions borrowed from phase transitions and critical phenomena. The qualitative model was transferred into five coupled, deterministic, nonlinear difference equations generating the dynamics of each variable in terms of other variables. (Such mathematical model is able to reproduce relevant features of psychotherapy processes.) Examples of parameter-dependent bifurcation diagrams are given. Beyond the illustrated similarities between simulated and empirical dynamics, the model has to be further developed, systematically tested by simulated experiments, and finally, compared to empirical data. In therapy processes of a variety of disorders, discontinuous trajectories of symptom changes have been identified. They occur in the treatment of patients with OCD following cognitive-behavioral group therapy. Time series analysis of data taken from about twenty OCD subjects revealed that a discontinuously shaped symptom reduction took place already before exposure (or, response) prevention in a good number of patients. The results support hypotheses from the theory of complex self-organizing systems (also discussed further within this study), postulating nonstationarity and critical instabilities during order transitions. The study delineates the usefulness of real-time monitoring procedures with high-frequency ratings, as given in terms of daily measurements in therapeutic routines, cf. Heinzel et al. (2014).

Developing further our theoretical arguments, in a normal diffusion mode, wherein there is no predictable directional data to be measured, the diffusion coefficient $D \xrightarrow{t \to \infty} E^2[W(t)]/t$, where $E^2[W(t)]$ is the expected variance of the fluctuations of the pertinent Wiener process $W(t)$; being always linear in $t$, this leads to a constant diffusion coefficient, or strictly speaking, after performing the above limiting operation, to an independency of $t$. Calling $E^1[W(t)]$ the expectation value of the process (Norris, 1998), one has $E^1[W(t)] = 0$ for any $W(t)$. Thus, one cannot define formally a key statistical quantity, namely the reduced variance (Gadomski and Ausloos, 2006), in the way introduced previously for $x$-s: $V_R(t) = \{E^2[W(t)] - (E^1[W(t)])^2\} / (E^1[W(t)])^2$ because of the prohibited division by zero. Whence, $E^1[W(t)] = 0$ always holds. (Bear in mind that our present, statistics-engaging concern, has been to get again the reduced quantity, $dx/x$; return to beginning of this section for recalling this.) The independence of $t$ demonstrates that behavioral actions are not dependent on time; these behaviors will consistently fail to show a preferential movement/tendency towards negative or positive activity in the subject. In this mode the behavior will not be indicative of progress or lack thereof.

If, instead, one introduces another process $P(t) \neq W(t)$ which could mimic some diffusion with a possibly small (constant) "depressurization" bias, or deflation drift, one would at once expect $E^1[P(t)] \neq 0$. This allows us, first, after introducing an anticipation rule, addressing the fluctuations, like $E^2[P(t+1)] = E^2[P(t)]$, to make some use of statistical independence requirements, as $E^2[P(t)] = (E^1[P(t)])^2$, and to firmly ascertain that $V_R(t) = 0$ ultimately applies. With this reduced variance, whether zero- or nonzero-in-value, one will be able to uncover the significance of subject behavior. The movement toward or away from progress can be seen once the variation of action, as having kept the form of reduced variance (Gadomski and Ausloos, 2006), is properly defined, which can only occur once the statistical independence of each action is asserted and when one may apply a simple (fluctuations' preservation addressing) anticipation rule within two consecutive periods of time.

It becomes a matter of simple inspection of statistical textbooks to read that the (above introduced) exponential, here so called Poisson distribution, such as $P(t) = \exp(-\lambda t)$, with some reciprocal scale parameter $\lambda > 0$, represents a tool for describing a consecutive passage from a state No. 1 (pressurized) to its depressurized counterpart (No. 2, i.e., a "deflated-balloon" state), with a transition probability $p_{1 \to 2}(t) = 1 - \exp(-\lambda t)$. This is due to the fact that $E^2[P(t)] = (E^1[P(t)])^2$ also applies here: the fluctuation in the (asymmetrically distributed) exponential process, i.e. $E^2[P(t)] = 1/\lambda^2$, equals its squared expectation value, since $E^1[P(t)] = 1/\lambda$ holds. The Poisson distribution furthers the idea that a subject's mental state, past experiences, and presumably genetic predisposition, are as significant to changes in actions and behavior as treatments are.

The well-known (and, seemingly deeply rooted) pitfall concerns here the fact that yet for some value of $k \cong 2.41$ also the (so called Pareto) power-law type distribution of the form $P(t) \propto t^{-(k+1)}$, thus with the necessary mathematical condition of $k > 2$, fairly obeys the same statistical independency requirements, i.e. also leading to $V_R(t) = 0$. (Note that this can virtually introduce some ambiguous and interpretation-addressing treatment of data, apparently obeying both statistics, albeit expected to have different origins of their emergence; cf. Izhikevich (2007).) One can be convinced about it after resolving the simple 2$^{nd}$ order algebraic equation $k^2 - 2k - 1 = 0$, giving the only positive root at $k \cong 2.41$ (characteristic exponent, see below), provided that another negative root ($k \cong -0.41$) is postponed. Then, as prejudiced above, the corresponding depressurization probability $p_{1 \to 2}(t) \cong 1 - (\lambda^{-1}/t^{2.41})$ differs distinctly, both quantitatively and qualitatively, by its temporal characteristic when compared to that of the exponential Poissonian nature. This observation provides a serious warning on utilizing the data that apparently conforms to either statistics. As a consequence, there appears a tremendous call for providing the reliability and publicity of the data (Guastello and Gregson, 2011) on which the empirically found distributions are based. This transparency should render the situation described in (Sijtsma, 2016) fairly avoidable, termed in the literature as "questionable research practices" (QRTs).

Self-organization, operating on the ground of scaling laws, is such a general theory, expressing the far-from-equilibrium subtleties of dissipative structures' (thermo)dynamics (Schiepek et al., 2015; Pincus, 2009), that may assist in providing a deeper, scientifically valuable understanding of the complex biopsychosocial processes involved in psychotherapy. To provide some foundation to this rather grand suggestion, the following rests upon more specific theoretical propositions stemming from self-organization theory (Pincus, 2009), especially those concerning deeper comprehension of self-organizing interpersonal processes in psychotherapy. To be specific, it has been shown that severe and persistent self-injurious behavior (SIB) is extremely difficult to elucidate. The subsequent study (Pinkus et al., 2014) used self-organization theory to investigate the possible relationship between SIB and changing levels of behavioral flexibility. Data consisted of categorical time-series of sequential behaviors from individuals with developmental disabilities and severe SIB. The method of orbital decomposition was used to analyze each series for measures of structure and entropy, a measure of behavioral disorder. In total, results indicated evidence for self-organization in complex behavior patterns. Next, series including SIB were on average more flexible than those without SIB; while, higher numbers of SIB events, coming from the perseveration, were associated with higher behavioral rigidity and structural disintegration. Finally, there was evidence that behavioral flexibility almost always shifts reliably after a discrete bout of SIB, either increasing or decreasing in complexity. Summing up, these results may provide a deeper and more theoretically grounded understanding of the function of SIB beyond the traditional behavioral paradigm involving simple stimulus-response or response-consequence relationships. Instead, some behaviors, such as SIB, may serve a resilience-making function as global regulators of behavioral flexibility and coherence.

One additional point to make is that it is now well accepted that nonlinear processes are, by and large, exponentially distributed or power-law distributed, Guastello and Gregson (2011). Pincus and Metton (2010) and Cerf (2015) also reported on cusp catastrophe processes; like other differentiable functions, they have a unique complex-exponential distribution of their own. The authors have brought together constructive work on new practical examples of methods and application built on nonlinear dynamics. They cover dynamics expressed by attractors, bifurcations, chaos, fractals (with their power laws), catastrophes, self-organization, and related issues in time series analysis, stationarity, modeling and hypothesis testing, probability, and experimental design. The analytic techniques discussed involve several variants of the fractal dimension, also several types of entropy, phase-space and state-space diagrams, recurrence analysis, spatial fractal analysis, oscillation functions, polynomial nonlinear regression, Markov chains with their ergodic properties, and finally, symbolic dynamics. Theoretical and methodological issues coming from nonlinear dynamical systems (NDS), of which algebraic and exponential behaviors are specific albeit important offsprings, may provide considerable advantage to health scientists as well as health care professionals. For instance, NDS methodologies and topics in health care share a focus upon the potentially complex interactions of biological, psychological and social factors over time, cf. Pincus and Metton (2010). Nevertheless, a number of challenges still remain in creating the necessary bridges in understanding to allow researchers to apply NDS techniques and to enable practitioners to use the resulting evidence to improve patient care. Thus, common concepts pertaining to self-organizing complex adaptive systems are outlined as a general approach to understanding resilience across biological, psychological, and social scales. To this end, four data analytic techniques from NDS have been compared and contrasted by Pincus and Metton (2010) with respect to the information they may contain about some common processes underlying resilience. These techniques are: time-series analysis, state-space grids, catastrophe modeling, and network modeling (Gadomski and Ausloos, 2006). Implications for health scientists and practitioners are discussed. It has been shown (Cerf, 2015) that a path not yet detected exists in the parameter space of the cusp catastrophe that constitutes a certain target-trajectory, along which psychological change may be achieved in a variety of situations by taking advantage of the protagonists' resistance. This mentioned target-trajectory offers: (i) conditions optimised in therapy with regard to the intrinsic limitations for the reduction of a patient's pathogenic agent, and in conflict with regard to the 'red-lines' of the protagonists, and (ii) the benefit of a step of rapid decrease in the potential barrier to change. Questions raised concern the benefit that a patient may obtain from performing his cognitive task in psychotherapy with minimal requirement for the reduction of his pathogen, and the role that a step of rapid decrease in a potential barrier may play in decision-making, especially when it comes to remove a conflict. The argument is well substantiated for psychoanalytic resistance, resting on principles and procedures described in numerous psychoanalytical texts. And, the most important message: the theory deals with scaling laws - power laws - rather than strict equalities, cf. Cerf (2015).

To recall the attention, the afore presented approach, but with the nonzero reduced variance, $V_R(t) \neq 0$, seems to be effective in qualitatively disclosing certain basic properties of mental viscoelastic network-like clusterings and interrelated dispersive (e.g., emotional) associations affecting, in a manner ideologically resembling those uncovered by (Gadomski and Ausloos, 2006), certain memory inconsistencies and dysfunctional (diffuse) effects, pointing foremost to tight or undense, chaotic type clusterings of (mental) associations, also being qualitatively reminiscent of either OCD (Heinzel et al., 2014) or even SIB (Pincus et al., 2014). Therefore, the zero-limit concerning reference way of $V_R(t) \to 0$ looks attractive and fully conceivable for carrying out quantitative psychological research (Schiepek et al., 2014) based on well-established monodisperse-like statistics.

## CONCLUSION AND PERSPECTIVE

To conclude, the <u>statistically based</u> approach presented in this communication, as compared, and in contrast, to its deterministic counterpart, based mainly on continuous time, may equally well be applicable for difficult multilevel psychological, psychotherapy engaging strategies, as proposed in (Gelfand and Engelhart, 2012; Osborne 2010), commencing here with a constant-drifted (asymmetric) diffusion process, $P(t)$, thus within a quasi-linear domain. A virtual control of hazard should be assured by some careful application of the negative $\Delta\pi$-type depressurization or deflation concerning factor. The multilevel interaction process of general exponential nature, such as the friction phenomenon, is also predisposed to possess some inherent psychological connotations. They can emerge in a way similar to how the interactions in the physical system are switched on and off (Lejeune, 1998; Bond and Guastello, 2013). The multilevel friction procedure can also support the overall argumentation line unveiled by the present communication, cf. Gadomski and Hładyszowski (2015). To make the approach more realistic and affordable for quantitative psychologists and cognition scientists, when uncovering it within a certain fractional time domain of primary, presumably therapeutic concern (West, 2015), the population based and capacity prone logistic Verhulstian model can be considered as a natural and useful alternative to its purely exponential counterpart that shows up either indefinite decays or grows (Mandel, 1995; Gadomski and Ausloos, 2006; Osborne, 2010).

As a last word, about distinguishing short and long term memory within DST, an extension to the notion of non-Markovianity (memory involvement), provided explicitly in terms of stochastic processes, can be invoked appropriately (Bazzani et al., 2003) via a Langevin type dynamical equation $dx/dt = -x + \xi^2(t)$ (thus, beyond Newtonian description of dynamics; see Cantley 2015), in which the additive quadratic, i.e. nonlinear (so-called) noisy, environment mimicking term, stands for the exponentially correlated Gaussian process (named Ornstein-Uhlenbeck process) with a zeroth expectation value, and a unit correlation time. (Note loosely that here also a signature of reduced variance, based on $dx/x$, implicitly emerges.) In (Łuczka et al., 1995), it has been demonstrated that the corresponding probability density $p(x,t)$ conforms to a drifted diffusion equation of Fokker-Planck type (i.e., obeying $V_R(t) \neq 0$), wherein both the drift and the diffusion coefficients, within the best approximation employed, do depend nonlinearly on the state variable $x = x(t)$. This is exactly the dynamic, albeit entropy-productive (Gadomski and Ausloos, 2006), manifestation of the fact that the system does not forget about its state - quite an essential landmark of each memory involving dynamical process (Fuliński et al., 1998). Within the Markovian approximation, thus memory free, the process becomes equipped with a certain multiplicative noise, Gaussian but uncorrelated (thus, in terms of independent fluctuations) – i.e., a signature of a bare diffusive fingerprint embodied in the system (Łuczka et al., 1995). For a (clinical) psychological treatment approach, this would imply that one could stimulate purposely a patient either with an added, and rather external $t$ – dependent (hazardous) perturbation, or try to introduce a typically uncontrolled variable, but in a controlled (confined) way, for example by a pharmacological, physical or physiological co-treatment, some internal dynamical responses of the organism under examination. Such a strategy may additionally help circumvent the QRPs pointed out in (Sijtsma, 2016), as well as to propose the so-called lay reader a companion that ensures certain further interpretations of the behavior expressed by an individual (Gelfand and Engelhart, 2012).

Finally, let us invoke a cognition science addressing study by (Favela, 2014), wherein he outlines the sand pile example, which involves aspects of both dynamics (DST proposed to be a methodology) and self-organized criticality as a theoretical, phase-transitive framework. (The overall model, involving the punctuated equilibria, is called the self organized criticality, abbreviated SOC). This example is easily adaptable in qualitative terms to a clinical model. It demonstrates that when one attempts to act on a subject, in a psychotherapeutic setting, the subject possesses an existing threshold. When this critical point is surpassed, one fails to influence the subject any further. Limiting much consideration to the goals of this communication, the importance of SOC cannot be further emphasized here. However, the clinical psychologist and/or cognitive scientist must be aware of this "critical point" when attempting to utilize DST, because the instability points that are taken into account in the dynamical systems theory cannot meet the sufficient certainty level that will ensure adherence to a subject's (human being's) individual critical point. Additionally, according to (Cantley, 2015), there is a mutual influence between patient and therapist. This is significant to the process of collecting data (Guastello and Gregson, 2011) for inputs into the DST or NDS related equations because, in reality, the process of data collection will be impacted by the influence of the patient on the therapist, and vice versa.

## ACKNOWLEDGMENT

This study is supported by the National Science Foundation – Philadelphia LSAMP (TC) and the UTP BS39/14 (AG).